\documentclass[twocolumn,showpacs,preprintnumbers,amsmath,amssymb]{revtex4}

\usepackage{graphicx}
\usepackage{dcolumn}
\usepackage{bm}
\begin{document}

\draft

\title{Stochastic Semi-Classical Description of Fusion at Near-Barrier Energies}
\author{Sakir Ayik$^{1}$, Bulent Yilmaz$^{2,3}$, and Denis Lacroix$^{3}$}
\address{$^{1}$Physics Department, Tennessee Technological University, Cookville, TN 38505, USA}
\address{$^{2}$Physics Department, Ankara University, 06100 Ankara, Turkey}
\address{$^{3}$Grand Acc\'el\'erateur National d'Ions Lourds (GANIL), CEA/DSM-CNRS/IN2P3, BP 55027, F-14076 Caen Cedex 5, France}

\date{\today}

\begin{abstract}
Fusion reactions of heavy ions are investigated by employing a simple stochastic semi-classical model which includes coupling between the relative motion and low frequency collective surface modes of colliding ions similarly to the quantal coupled-channels description. The quantal effect enters into the calculation through the initial zero-point fluctuations of the surface vibrations. A good agreement with results of coupled-channels calculations as well as experimental data is obtained for fusion cross sections of nickel isotopes. The internal excitations in non-fusing events as well as the fusion time are investigated.  
\end{abstract}

\pacs{24.10.Lx, 25.60.Pj, 25.70.Jj}

\maketitle

\section{Introduction}
 
Over the past years various models with different levels of approximation have been developed to explain the enhancement of sub-barrier fusion cross sections close to the Coulomb barrier energy. The description of fusion reactions is generally realized by phenomenological models such as barrier penetration models (see ref.  \cite{Balantekin} for a review) or more sophisticated approaches like coupled-channels approach \cite{Brown,Landowne,Hagino}. These models provide some qualitative as well as quantitative explanations to the enhancement of the cross sections as a result of coupling to selective collective modes. In such macroscopic approaches, one of the main ingredients is the nuclear interaction potential which can be taken as the Bass potential \cite{Bass}, the proximity potential \cite{Blocki,Randrup}, or the double-folding potential \cite{Satchler}. Some microscopic approaches equipped with full quantum treatment such as Time Dependent Hartree-Fock model are also used to give quantitative explanations of nuclear fusion reactions by providing a connection between the macroscopic and microscopic phenomena \cite{Kim,Simenel,Umar,Washiyama,Simenel2}. 

In this paper, we investigate fusion reactions of heavy-ions by employing a simple stochastic semi-classical model. In this model, the coupling between the relative motion and low frequency collective surface modes is incorporated into the description in a manner similar to quantal coupled-channels calculations. Quantal effects enter into the calculation through the initial zero-point fluctuations of the surface modes. In the calculations, we include the coupling of relative motion with the low-lying quadrupole and octupole surface modes. The zero-point fluctuations of surface vibrations are simulated in a stochastic approximation by generating an ensemble of trajectories in accordance with the Gaussian distribution of zero-point fluctuations. The zero-point quantum fluctuations of surface modes lead to barrier fluctuations, which enhance the fusion cross-section at near and sub-barrier energies. This stochastic semi-classical model has been proposed in ref. \cite{Esbensen} (see also \cite{Dasso} for a detailed review). However, only a few applications of the model have been carried out so far \cite{Guidry,Galetti,DassoDonangelo}. This is probably due to insufficient computation power at the time it was proposed. In this work, employing this stochastic model, we investigate fusion reactions of nickel isotopes at near-barrier and sub-barrier energies. We compare our results with the experimental data as well as the quantal coupled-channels calculations with the same input parameters. In section two, the model is briefly described. In section three, calculations of the fusion cross-sections of nickel isotopes are presented. In section four, further applications of the approach are illustrated and conclusions are given in section five.

\section{Stochastic Semi-Classical Model}

In order to describe heavy-ion collisions and fusion at near-barrier and sub-barrier energies, we follow the idea originally proposed by Esbensen et al.
\cite{Esbensen}. In this semi-classical model, heavy-ion collisions are described by a Hamiltonian in which the relative motion is coupled to a number of low-lying collective modes (surface vibrations) of the colliding ions. Treating the surface vibrations in harmonic approximation, the semi-classical Hamiltonian for the colliding ions is given as, 
\begin{eqnarray}
 \label{e1}
H&=& \frac{P^2}{2\mu}+\frac{l(l+1)\hbar^2}{2\mu R^2}+V_C(R)+V_N(R,\Omega,\alpha_{i\lambda})\nonumber\\
&&+\sum_{i=1}^2\sum_{\lambda=0}^{N-1}\left[\frac{\Pi_{i\lambda}^2}{2D_{i\lambda}}+\frac{1}{2}C_{i\lambda}
\alpha_{i\lambda}^2\right],
\end{eqnarray}
where $R$ represents the relative distance between two centers of mass of the colliding nuclei and $P$ is the corresponding relative momentum. In this expression, the first and second terms are the radial kinetic energy and the rotational kinetic energy with orbital angular momentum $l$. The quantities $V_C(R)$ and $V_N(R,\Omega,\alpha_{i\lambda})$ represent the Coulomb potential energy and the nuclear interaction potential. The parameters set $\Omega=\{\Omega_1,\Omega_2,\Omega_3\}$ in the nuclear potential describes rotation angles of the vibration axes of the nuclei, which are specified below. The last term in Eq. (\ref{e1}) is the 
Hamiltonian for 2N harmonic oscillators corresponding to the vibrational modes $(\lambda=0,...,N-1)$ of projectile and target ions $(i=1,2)$. The quantities $\alpha_{i\lambda}$, $\Pi_{i\lambda}$, $D_{i\lambda}$, and $C_{i\lambda}$ indicate variables for the vibrational modes, that correspond to the deformation variables, the corresponding momenta, inertia parameters, and spring constants, respectively. The spring constants and the inertia parameters of harmonic oscillators are determined in terms of the deformation parameters $\beta_{i\lambda}$ and the excitation energies $E_{i\lambda}^\star$ of the modes according to
$C_{i\lambda}=E_{i\lambda}^\star/2\beta_{i\lambda}^2$ and $D_{i\lambda}=\hbar^2/2E_{i\lambda}^\star\beta_{i\lambda}^2$, 
respectively \cite{DassoDonangelo,Esbensen4}. In the ground state, the variances of vibrational variables and the variances of the corresponding momentum are expressed according to $\sigma_{\alpha_{i\lambda}}=\beta_{i\lambda}$ and $\sigma_{\Pi_{i\lambda}}=\hbar/2\beta_{i\lambda}$. 

The classical equations of motion for the relative distance and the vibrational variables are given by, 
\begin{eqnarray}
 \label{sol}
 \frac{d R}{d t}&=&\frac{P}{\mu},\nonumber\\
 \frac{d P}{d t}&=&-\frac{d V_C(R)}{d R}
 -\frac{\partial V_N(R,\Omega,\alpha_{i\lambda})}{\partial R}+\frac{l(l+1)\hbar^2}{\mu R^3},\nonumber\\
 \frac{d \alpha_{i\lambda}}{d t}&=&\frac{\Pi_{i\lambda}}{D_{i\lambda}},\nonumber\\
 \frac{d \Pi_{i\lambda}}{d t}&=&-\frac{\partial V_N(R,\Omega,\alpha_{i\lambda})}{\partial\alpha_{i\lambda}}
 -C_{i\lambda}\alpha_{i\lambda}. 
\end{eqnarray}
In the model, a dissipation of the relative energy occurs due to excitations of the surface modes. An important dissipation mechanism due to nucleon exchange between projectile and target nuclei \cite{Randrup2,Feldmeier} is neglected here. However, at low bombarding energies, the mechanisms related to excitations of the surface modes dominate compared to the nucleon exchange mechanism (see Figure 1 of ref. \cite{Denisov}). Consequently, the model provides a deterministic description of the average properties of collision dynamics at low energies.

Due to a short de Broglie wavelength, the classical approximation works well for relative motion at near-barrier energies where the effect of tunneling is small compared to that of the surface excitations. However, as a result of a few phonon excitations during the collision process, the dynamics of surface vibrations is far from the classical limit, and should be treated in a quantal framework. A standard description is provided by the quantal coupled-channels calculations. Here, instead of the standard coupled-channels description, we include the quantal fluctuations of the surface modes by incorporating the initial zero-point fluctuations of the vibrational modes. We can determine the phase space distribution function $F(\alpha,\Pi)$ of a harmonic oscillator ground state, by taking the Wigner transform of the ground state wave function to find, 
\begin{eqnarray}
 \label{wig}
 F(\alpha,\Pi)=\frac{1}{2\pi\sigma_\alpha\sigma_\Pi}
 \exp\left(-\frac{\alpha^2}{2\sigma^2_\alpha}-\frac{\Pi^2}{2\sigma^2_\Pi}\right),
\end{eqnarray}
where $\alpha=\alpha_{i\lambda}$ and $\Pi=\Pi_{i\lambda}$ are the variances of coordinate and momentum distributions for each vibrational mode. The quantal zero-point fluctuations of the vibrational modes are incorporated in a stochastic manner. An ensemble of trajectories is generated by solving the classical equations of motion with initial conditions $\alpha_{i\lambda}(0)$ and $\Pi_{i\lambda}(0)$, that are randomly selected according to the corresponding Wigner distribution $F(\alpha,\Pi)$. Once the ensemble of trajectories is generated, different observables are calculated by averaging over the ensemble.
  
The following approximation for the Coulomb potential is employed, 
\begin{eqnarray}
 \label{e2}
 V_C(R)=\left\{
 \begin{array}{c}
 \frac{Z_1Z_2e^2}{R} \qquad\qquad\qquad ~~ R>R_C \\ 
 \frac{Z_1Z_2e^2}{R_C}\left(\frac{3}{2}-\frac{1}{2}\frac{R^2}{R_C^2}\right) \quad R<R_C
 \end{array}
 \right. ,
\end{eqnarray}
where $R_C=R_1+R_2$ is the sum of the equivalent sharp radii \cite{Birkelund,Hasse}. The nuclear part of the interaction is computed using the double-folding potential as,
\begin{eqnarray}
 \label{e3}
 V_N(R,\Omega,\alpha_{i\lambda})&=&\int\rho_1(\vec{r}_1,\Omega_1,\alpha_{1\lambda})
 \rho_2(\vec{r}_2,\Omega_2,\alpha_{2\lambda})\nonumber\\
 &&\times V_{NN}(\vec{R}-\vec{r}_1+\vec{r}_2)d^3r_1d^3r_2.
\end{eqnarray}
Fig. 1 illustrates the geometry of two colliding ions in which $x_1, y_1, z_1$ and $x_2, y_2, z_2$ denote two sets of coordinate systems with fixed orientations and origins that are attached to the centers of the nuclei. The red lines indicate the vibration directions. The angles between position vectors $\vec{r}_1$, $\vec{r}_2$ and the axes $z_{1,2}$ are indicated by $\theta_1$ and 
$\theta_2$. The angles $\theta'_1$ and $\theta'_2$ represent the angles between position vectors $\vec{r}_1$, $\vec{r}_2$ and the vibration directions of the nuclei, respectively. The nuclear folding potential, in addition to the relative position $R$ and the vibrational variables $\alpha_{i\lambda}$, also depends on three independent angles $\Omega=\{\Omega_1,\Omega_2,\Omega_3\}$ which specify the relative orientation of      
vibration directions of the nuclei. 
\begin{figure}[htp]
\includegraphics[width=3.4in]{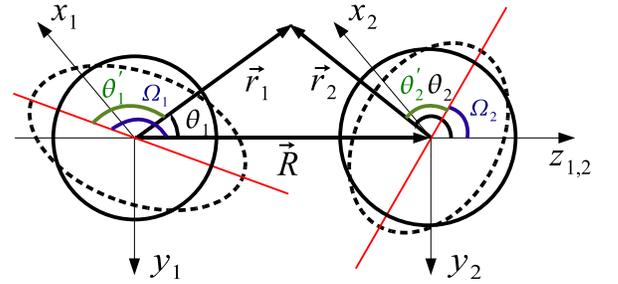}
\label{figtwonuc}\caption{(color online) A schematic view of nuclei with quadrupole vibrations. The red lines indicate the vibration directions.}
\end{figure}
The angles $\Omega_1$ and $\Omega_2$ are defined as the rotation angles of the vibration directions about $x_1$ and $x_2$ axes, respectively. These two angles define orientations within the same plane ($x_{1,2}=0$), and in order to cover all possible orientation configurations, one more angle $\Omega_3$, is needed to rotate the oscillation axis of one of the nuclei about the corresponding $y$ or $z$ axis to account for the off-plane orientations. For convenience, a rotation about the $y$ axis is considered. The $\Omega_3$ rotation is not indicated in the figure. We approximate the nuclear density distributions by two-parameter Fermi functions, which can be conveniently expressed in terms of the angles $\theta'_1$ and $\theta'_2$ as  
\begin{eqnarray}
 \label{e4}
 \rho_i(\vec{r}_i,\Omega_i,\alpha_{i\lambda})&\equiv&\rho_i(r_i,\theta'_i,\alpha_{i\lambda})\nonumber\\
 &=&\frac{\rho_{0i}}{1+\exp\left[(r_i-R_i(\theta'_i,\alpha_{i\lambda}))/a_i\right]},
\end{eqnarray}
where $\rho_{0i}$ is a normalization constant, $a_i$ is the diffuseness parameter and $R_i(\theta'_i,\alpha_{i\lambda})$ denotes the deformed nuclear radius of each nucleus. For small amplitude vibrations, this quantity is expanded in terms of spherical harmonics, 
\begin{eqnarray}
 \label{e5}
 R_i(\theta'_i,\alpha_{i\lambda})&=& 
 R_{0i}\left(1+\sum_\lambda\alpha_{i\lambda}Y_{\lambda 0}(\theta'_i)\right)\nonumber\\
&=& R_{0i}\left(1+\sum_\lambda\alpha_{i\lambda}\sqrt{\frac{2\lambda+1}{\pi}}P_{\lambda}(\cos\theta'_i)\right).
\end{eqnarray} 
From the geometry displayed in Fig. 1, it is possible to deduce the following relations between angles, 
\begin{eqnarray}
\label{e55}
 \cos\theta'_1&=&\cos\Omega_1\cos\theta_1 - \sin\Omega_1\sin\theta_1\sin\phi_1,\\
 \nonumber\\
 \cos\theta'_2&=&\cos\Omega_2(\cos\Omega_3\cos\theta_2-\sin\Omega_3\sin\theta_2\cos\phi_2)\nonumber\\
   && -\sin\Omega_2\sin\theta_2\sin\phi_2.\label{e56}
\end{eqnarray}
The normalization constants $\rho_{0i}$ in Eq. (\ref{e4}) are obtained from the equation, 
\begin{eqnarray}
\label{e6}
 \int\rho_i(\vec{r}_i,\Omega_i,\alpha_{i\lambda})d^3r_i=A_i,
\end{eqnarray}
where $A_i$ is the mass number of the $i^\text{th}$ nucleus. For the equivalent sharp radii of the spherical nuclei, we take the values given by $R_{0i}=1.31A_i^{1/3}-0.84$ fm. 

A global description of the nucleus-nucleus potential can be achieved via the folding potential Eq. (\ref{e3}) by considering a zero-range nucleon-nucleon interaction,
\begin{eqnarray}
 \label{e7}
 V_{NN}(\vec{r})=V_0\delta(\vec{r}),
\end{eqnarray}
which is equivalent to a finite-range nucleon-nucleon interaction \cite{Chamon,Ribeiro,Chamon2,Gasques}. Then, the nuclear potential Eq. (\ref{e3}) becomes
\begin{eqnarray}
 \label{e8}
V_N(R,\Omega,\alpha_{i\lambda})=\qquad\qquad\qquad\qquad\qquad\qquad\qquad\quad\nonumber\\
V_0\int\rho_1(r_1,\theta'_1,\alpha_{1\lambda})
 \rho_2(r_2,\theta'_2,\alpha_{2\lambda})r_1^2dr_1d(\cos\theta_1)d\phi_1,
\end{eqnarray}
where $\theta'_1$ and $\theta'_2$ are given by Eq. (\ref{e55}) and Eq. (\ref{e56}), respectively with 
\begin{eqnarray}
\label{e9}
 \cos\theta_2&=&(r_1\cos\theta_1-R)/r_2,\\
 r_2&=&\sqrt{r_1^2+R^2-2r_1R\cos\theta_1},\\
 \phi_2&=&\phi_1.
\end{eqnarray}
In order to simplify the numerical simulations, a further approximation is introduced. For small amplitude vibrations, Taylor expansions 
of the nuclear densities and the nuclear potential are introduced to the first order around $\alpha_{i\lambda}=0$ to give,
\begin{eqnarray}
 \label{e105}
 \rho_i(\vec{r}_i,\Omega_i,\alpha_{i\lambda})
&\approx& \rho_i(\vec{r}_i,\Omega_i,0)\qquad\qquad\qquad\qquad\qquad\quad\nonumber\\
&+&\sum_{\lambda}\alpha_{i\lambda}\left[\frac{\partial}{\partial\alpha_{i\lambda}}
\rho_i(\vec{r}_i,\Omega_i,\alpha_{i\lambda})|_{\forall\alpha=0}\right],
\end{eqnarray}
\begin{eqnarray}
 \label{e11}
 V_N(R,\Omega,\alpha_{i\lambda})&\approx& V_N(R,\Omega,0)\qquad\qquad\qquad\qquad\qquad\quad\nonumber\\
&+&\sum_{i,\lambda}\alpha_{i\lambda}\left[\frac{\partial}{\partial\alpha_{i\lambda}}
V_N(R,\Omega,\alpha_{i\lambda})|_{\forall\alpha=0}\right],
\end{eqnarray}
respectively. 

Fig. 2 shows examples of potential energies in head-on collisions of two $^{64}$Ni nuclei for different orientations and deformations as a function of the relative distance. 
In this figure, for simplicity, only the quadrupole vibrations of both nuclei in two different orientations are shown. One of them corresponds to the case when vibrations are along the direction of the relative motion indicated by blue curves. 
\begin{figure}[htp]
\includegraphics[width=3.4in]{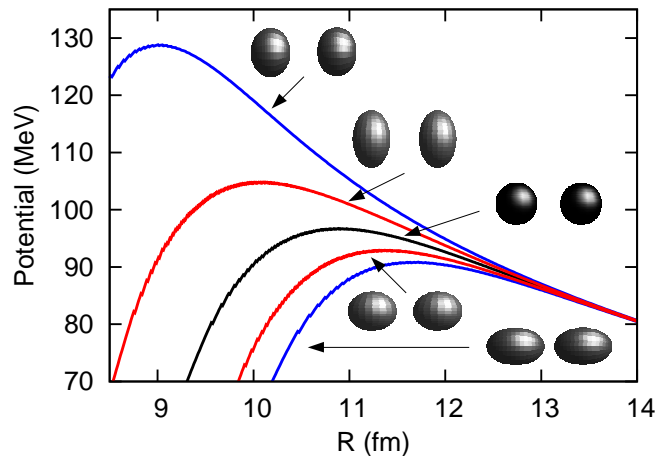}
\label{figPot}\caption{(color online) The s-wave potential barriers of $^{64}$Ni + $^{64}$Ni system are plotted versus center to center distance for different quadrupole deformations and orientations. The blue curves represent the oblate (upper curve) and prolate (lower curve) deformations along the relative motion. The red curves stand for the prolate (upper curve) and oblate (lower curve) deformations that are perpendicular to the relative motion direction. The bare potential with no deformation is indicated by the black color. The potentials are computed by summing the Coulomb $V_C$ and nuclear $V_N$ contributions with deformation variables equal to their variances, $|\alpha_{12}|=|\alpha_{22}|=\beta_2=0.215$. The zero-range potential strength in Eq. (\ref{e7}) is taken as $V_0=-456$ MeVfm$^3$ \cite{Chamon}.}
\end{figure}  
In the following, this orientation is referred to as the ZZ configuration. 
The other orientation corresponds to the case when vibrations are perpendicular to the direction of the relative motion indicated by red curves.
We refer to this orientation as the YY configuration. It is observed that in the ZZ configuration, the difference between the potential barriers for oblate and prolate deformations is larger than the difference in the YY configuration. Any orientation in between these two configurations leads to a difference in barrier heights, which is smaller than that of the ZZ configuration and larger than that of the YY configuration. Hence, the ZZ and YY configurations represent the extreme states of barrier fluctuations due to surface vibrations. When the off-plane orientations are included 
($\Omega_3 \neq 0$), the difference in barrier fluctuations becomes minimum for the configuration where vibrations are perpendicular to the direction of the relative motion and perpendicular to each other. This orientation is referred to as the XY configuration, which is not included in Fig. 2. It is clear that different orientations of the nuclear surface vibrations have a very large influence on the fusion barrier fluctuations. Therefore, we need to calculate any observables by averaging over all possible relative orientations of the vibration directions. We carry out this averaging by sampling all three angles, \{$\Omega_1$, $\Omega_2$, $\Omega_3$\}, from a uniform distribution in the interval [$0$,$2\pi$]. 

\section{Fusion Cross-Sections}

Coupled-channels calculations are often employed for describing fusion cross-sections at sub-barrier energies. These investigations indicate that low-lying surface modes such as $2^+$ and $3^-$ make the dominant contribution to sub-barrier cross-sections \cite{Esbensen2,Esbensen3,Beckerman2}. Retaining only these two modes, we carry out stochastic simulations to describe the fusion process of Nickel isotopes. In order to compare our results with that of coupled-channels calculations of Nobre et al. \cite{Nobre}, we adopt the same parameters as in that reference. It is important to note that none of these parameters are adjustable. The quadrupole ($\lambda=2$) and octupole ($\lambda=3$) deformation parameters are 
$\beta_2=0.215$, $\beta_3=0.263$ for $^{64}$Ni, and $\beta_2=0.205$, $\beta_3=0.235$ for $^{58}$Ni. The excitation energies are 
$E_2^\star=1.35$ MeV, $E_3^\star=3.56$ MeV for $^{64}$Ni, and $E_2^\star=1.45$ MeV, $E_3^\star=4.48$ MeV for $^{58}$Ni. The zero-range potential strength and the diffuseness parameter are $V_0=-456$ MeVfm$^3$ and $a=0.56$, respectively \cite{Chamon}. For simplicity, we consider that the quadrupole and octupole vibrations of each nucleus are aligned in the same direction. 
\begin{figure}[htp]
\includegraphics[width=3.4in]{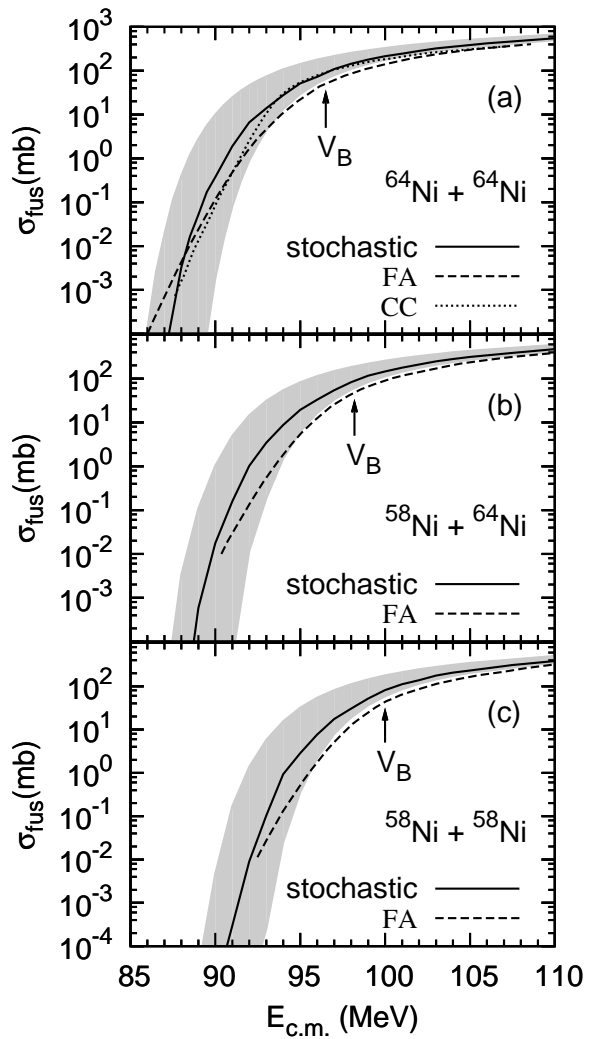}
\label{figcros1}\caption{Fusion cross sections of $^{64,58}$Ni + $^{64,58}$Ni systems calculated with the coupled-channels model (CC) and the Frozen Approximation model (FA) of Nobre et al. \cite{Nobre} are compared with our results (the solid line). Barrier heights are indicated in the figures. The shaded areas are explained in the text.}
\end{figure}

We calculate the fusion cross section using the standard expression,
\begin{eqnarray}
 \label{cross}
 \sigma_{fus}(E)=\frac{\pi\hbar^2}{2\mu E}\sum_{l=0}^{l_{max}}(2l+1)P_l(E),
\end{eqnarray}
where $E$ and $P_l(E)$ represent the incident center of mass bombarding energy and the partial transmission probabilities, respectively. 
Fig. 3 shows the results of the stochastic calculations and comparisons with the coupled-channels calculations. The solid line corresponds to the cross section obtained by averaging over all orientations. The shaded area in the figure illustrates the cross section fluctuations due to the effect of different vibration orientations. The upper (lower) boundary of the shaded area corresponds to the cross section when vibrations are along (perpendicular to) the relative motion direction which gives rise to maximum (minimum) cross-section due to the largest (smallest) barrier fluctuations. The coupling to surface vibrations increases the fusion cross-sections at sub-barrier energies, whereas it decreases the cross-sections at over-barrier energies. This effect is further enhanced due to the orientation configurations which can be easily visualized by looking at Fig. 3. At very low bombarding energies, the stochastic fusion cross-sections tend to approach the ZZ configuration limit since the barrier fluctuations are larger in this case. On the other hand, at over-barrier energies, the fusion cross-sections tend to approach the XY configuration limit since the transmission is reduced by very large barrier heights appearing in the ZZ configuration. The reduction in the ZZ configuration is eventually increasing the weight of events in the XY configuration. In this figure, we also compared our results with the coupled-channels calculations (CC) and the frozen approximation (FA) of Nobre et al. \cite{Nobre} which is an approximation to the coupled-channels calculations. Even though same parameters are used in both calculations, there are some minor differences. While they used the double folding potential in their CC calculations, a frozen density approximation is further assumed. Both calculations, CC and FA, are performed with a parabolic approximation for the potential barriers with effective curvatures. The stochastic approach does not have these drawbacks, which are important at low sub-barrier energies. Furthermore, in the stochastic description, the effects of vibration direction orientation of the nuclei is incorporated in a natural way. On the other hand, the stochastic model does not include the effect of quantum tunneling. Consequently, the stochastic approach provides a good description at near-barrier energies where barrier fluctuations due to the surface vibrations provide dominant contributions to the transmission probability. In Fig. 3, it is seen that the stochastic semi-classical description provides a good approximation to the quantal CC model and its approximate FA version for nickel isotopes fusion at near-barrier energies. In Fig. 4, the fusion excitation functions obtained by the stochastic semi-classical model are compared with three different data sets. Our results are in good agreement with the data near and below the barrier. The model over-estimates the experimental cross-sections at over-barrier energies due to the following reasons. First, the nuclear potential that we employ is energy independent. Energy dependence of nuclear potential can have a significant effect on fusion mechanism above the barrier. Actually, there are contradictory claims for the energy dependence of nuclear potentials. In ref. \cite{Chamon,Ribeiro,Chamon2}, it is claimed that, at near-barrier energies, the energy dependence vanishes while at higher energies it becomes important. On the other hand, in some microscopic calculations, an opposite behavior is found \cite{Washiyama}. Second, the folding potential that is employed here has a deep minimum as a result of the zero-range interaction. It has been shown that a repulsive core within the M3Y nucleon-nucleon interaction gives rise to shallower, more realistic nuclear potentials \cite{Misicu}. The maximum angular momentum $l_{max}$, up to which the partial transmissions are summed over in Eq. (\ref{cross}), is determined by the fact that the potential pocket disappears. For deep potentials, $l_{max}$ is larger than that for shallow potentials, which increases the over-barrier cross sections. We also believe that the lack of dissipation in the relative motion as well as in surface vibrations is the third reason for the over-estimation of data above the barrier. Dissipation mechanism can be easily incorporated into the stochastic description. Nevertheless, it is not considered in this study. Our task in this work is to test the stochastic semi-classical model by carrying out simulations of fusion mechanism and compare the results with the quantal coupled-channels calculations. Therefore, we avoid employing a complicated nuclear potential.
\begin{figure}[htp]
\includegraphics[width=3.4in]{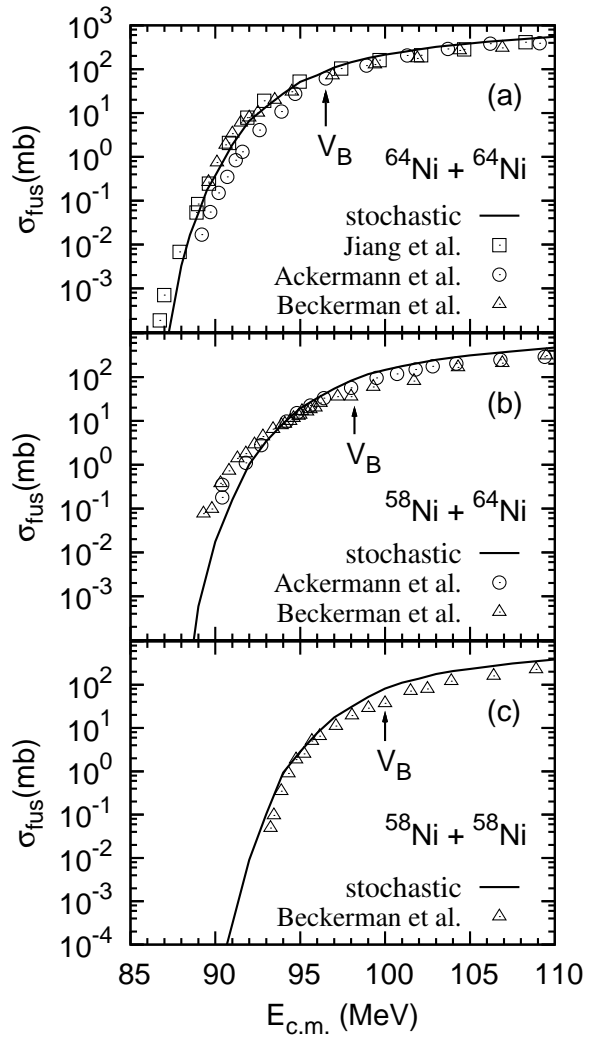}
\label{figcros2}\caption{The fusion cross sections of $^{64,58}$Ni + $^{64,58}$Ni systems calculated with the stochastic zero-point model are compared with the data of ref. \cite{Jiang,Ackermann,Beckerman}. Barrier heights are indicated in the figures.}
\end{figure}

\section{Discussions}

The stochastic semi-classical model provides a simple framework not only for fusion cross-sections but also to evaluate some relevant observables such as the time distribution for fusion and the kinetic energy distribution of non-fusion events which can hardly be accessed in a fully quantal framework. Here, we restrict the study to collisions between $^{64}$Ni nuclei. 
\begin{figure}[htp]
\includegraphics[width=3.4in]{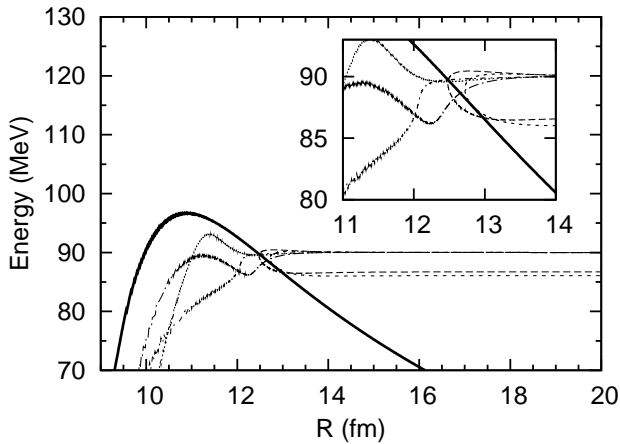}
\label{figtraj}\caption{The center of mass energies for five sample events, with incident energy of 90 MeV, are plotted versus the relative distance of two  $^{64}$Ni nuclei. The center of mass energy is the sum of the first four terms in Eq. (\ref{e1}), hence it is the energy of the relative motion. The angular momentum is set to zero. The s-wave potential barrier is shown for the bare case where there is no coupling to the surface modes (thick line).}
\end{figure}
\begin{figure}[htp]
\includegraphics[width=3.4in]{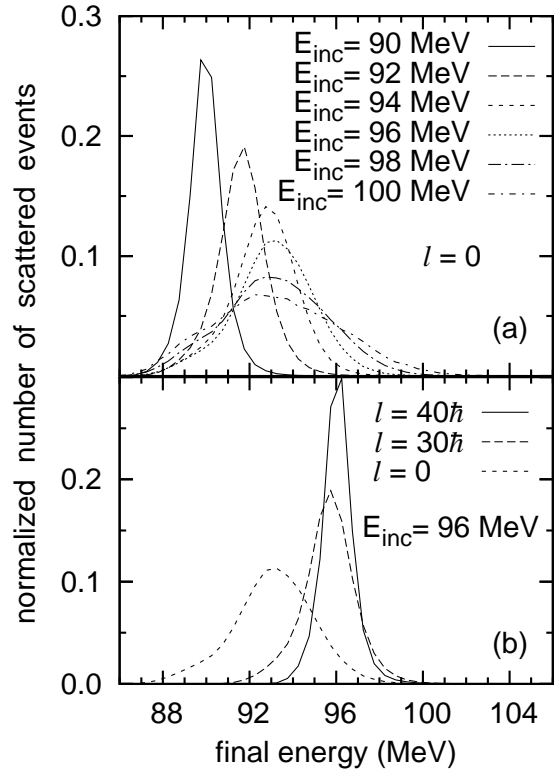}
\label{figdise}\caption{The distribution of the final energy of non-fusing events in collisions of $^{64}$Ni nuclei is plotted for different incident center of mass energies (a) and different angular momenta (b). For each plot the distribution function is normalized with the total number of scattered (no-fusion) events.}
\end{figure}
Fig. 5 shows five illustrative events as a function of relative distance of $^{64}$Ni ions for head-on collisions at bombarding energy of 90 MeV. In this figure, the thick line indicates the bare potential. Even though the bombarding energy is below the bare barrier, as a result of barrier fluctuations, three of the five events end up fusing, while two events after inelastic collision re-separate. During inelastic collisions, part of the incident energy is dissipated by excitations of the surface vibrations. Using the ensemble of events generated for description of the collision process, we can calculate the final kinetic energy distributions of the non-fusion events. 
\begin{figure}[htp]
\includegraphics[width=3.4in]{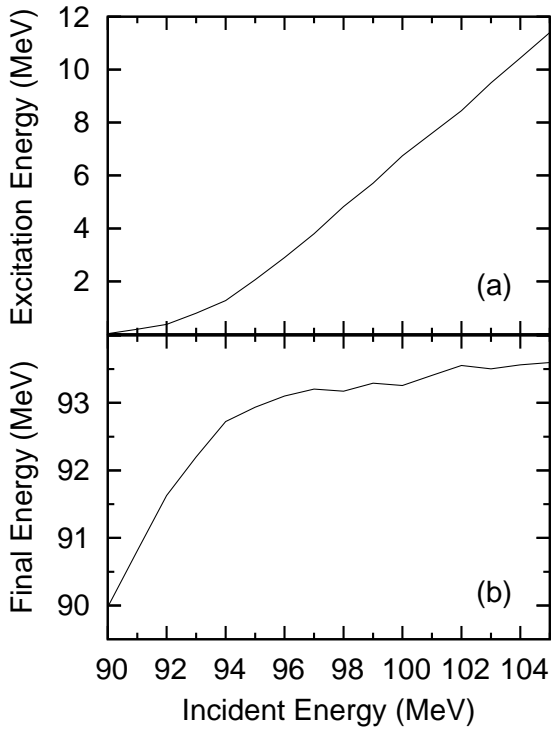}
\label{figfin}\caption{The average values of the total excitation energy (a) and the final energy (b) in head-on collisions of $^{64}$Ni nuclei are plotted as a function of the incident center of mass energy. The excitation energy is given by $E^\star= E_\text{inc}-E_\text{final}$.}
\end{figure}
The upper panel (a) of Fig. 6 shows final kinetic energy distributions of non-fusion events at different incident energies in head-on collisions of $^{64}$Ni + $^{64}$Ni systems, while the lower panel (b) presents the final kinetic energy distributions for different relative angular momenta at an incident energy of 96 MeV. As it is expected, the mean final energy increases with the incident energy as well as with the angular momentum. The variances of the distributions increase as incident energy increases and angular momentum decreases. Indeed, in this case, the rates of non-fusion events decrease eventually increasing the uncertainty of the final energy. 
Fig. 7 illustrates the average value of the final energy of the relative motion of non-fusion events. It is observed that, at very low incident bombarding energies, almost no energy is transferred to surface excitations. 
As the energy increases, the amount of dissipated energy into the surface modes increases. After the energy exceeds the fusion barrier height at $V_B=96.5$ MeV, the mean final energy is almost constant indicating that the total excitation energy $E^\star$ is almost linearly increasing with the incident energy. In the simulations of the trajectories, we take the initial relative distance as $R=20$ fm. We call the event as a fusion event, if the trajectory evolves all the way until the separation distance reaches to $R=5$ fm. For fusion events, we define the time it takes to travel from $R=20$ fm to $R=5$ fm as the fusion time. The upper panel (a) of Fig. 8 indicates the distribution of fusion times at different incident energies in central collisions of nickel ions, while the lower panel (b) shows the fusion time distributions for three different orbital angular momenta at incident energy of 96 MeV. Again, as we expect, the mean fusion time increases with increasing angular momentum and decreasing incident energy.

\section{Conclusion}

Employing a stochastic semi-classical model for low-energy heavy-ion collisions, which was proposed originally by Esbensen et al. \cite{Esbensen}, we carry out simulations for describing fusion and some other properties of the collision of nickel isotopes at sub-barrier and near-barrier energies. As known from coupled-channels calculations, dominant effects for describing sub-barrier fusion arise from the coupling of the relative motion with the low-lying surface vibrations. Therefore, in analogy to coupled-channels calculations, we consider the coupling of the relative motion with low-lying collective surface vibrations. Since the de Broglie wavelength is very short, a classical treatment provides a good approximation for the relative motion. However, the qauntal aspects of surface vibrations play a dominant role in the sub-barrier fusion mechanism. In the present approach, we incorporate quantal zero-point fluctuations of the surface vibrations in a stochastic approximation. In the applications presented here, only the quadrupole and octupole vibrations of the projectile and target nickel ions are included. An ensemble of trajectories of the relative motion are generated by picking the initial conditions of surface vibrations according to the quantal zero-point fluctuations of their ground states. Some observables such as fusion cross-sections and final kinetic energy distributions of non-fusion inelastic collisions have been estimated by averaging over the ensemble of trajectories. The simple stochastic semi-classical approach provides a surprisingly good agreement with quantal coupled-channels calculations as well as the experimental data for fusion cross-sections of nickel isotopes at sub-barrier and near-barrier energies. Of course, barrier penetration is not included in the description. However, at near-barrier energies, the dominant effects on fusion arise from the barrier fluctuations, which are well accounted for by the stochastic approach.  
\begin{figure}[htp]
\includegraphics[width=3.4in]{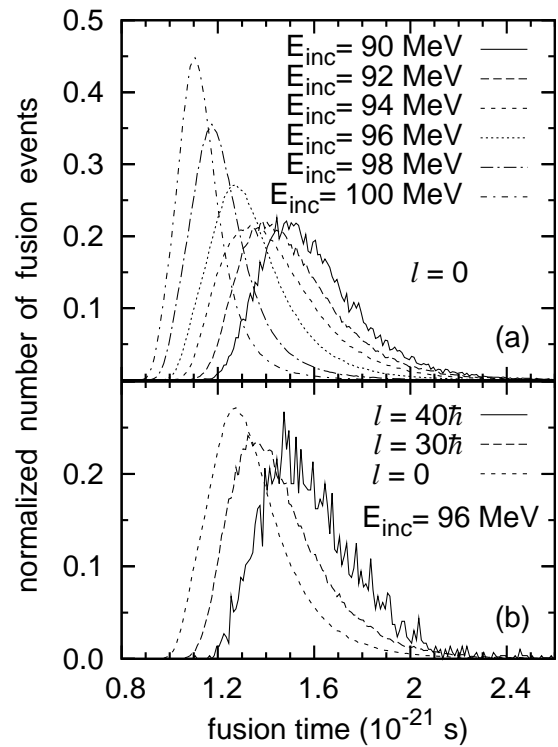}
\label{figdist}\caption{The distributions of the fusion time in collisions of $^{64}$Ni nuclei are plotted for different incident center of mass energies (a) and different angular momenta (b). For each plot the distribution function is normalized with the total number of fusion events.}
\end{figure}
 
\begin{acknowledgements} 
S.A. and B.Y. gratefully acknowledge GANIL, and also S.A. and D.L. gratefully acknowledge TUBITAK and Middle East Technical University for their support and warm hospitality extended them during their visits. This work is supported in part by the US DOE Grant No. DE-FG05-89ER40530.
\end{acknowledgements}


\begin{thebibliography}{99}
\bibitem{Balantekin} A. B. Balantekin and N. Takigawa, Rev. Mod. Phys. 70, 77 (1998).
\bibitem{Brown} M. J. Rhoades-Brown and M. Prakash, Phys. Rev. Lett. 53, 333 (1984).
\bibitem{Landowne} S. Landowne and S. C. Pieper, Phys. Rev. C 29, 1352 (1984).
\bibitem{Hagino} K. Hagino, N. Rowley, and A. T. Kruppa, Comput. Phys. Commun. 123, 143 (1999).
\bibitem{Bass} R. Bass, Nucl. Phys. A 231, 45 (1974).
\bibitem{Blocki} J. Blocki, J. Randrup, W. J. Swiatecki, and C. F. Tsang, Ann. Phys. (NY) 105, 427 (1977).
\bibitem{Randrup} J. Randrup and J. S. Vaagen, Phys. Lett. B77, 170 (1978).
\bibitem{Satchler} G. R. Satchler and W. G. Love, Phys. Rep. 55, 183 (1979).
\bibitem{Kim} K.-H. Kim, T. Otsuka and P. Bonche, J. Phys. G 23, 1267 (1997).
\bibitem{Simenel} C. Simenel, Ph. Chomaz, and G. de France, Phys. Rev. Lett. 93, 102701 (2004).
\bibitem{Umar} A. S. Umar and V. E. Oberacker, Phys. Rev. C 77, 064605 (2008); Phys. Rev. C 74, 061601(R) (2006); 
Phys. Rev. C 74, 024606 (2006).
\bibitem{Washiyama} K. Washiyama and D. Lacroix, Phys. Rev. C 78, 024610 (2008).
\bibitem{Simenel2} C. Simenel, B. Avez, and D. Lacroix, arXiv:0806.2714 (2009). 
\bibitem{Esbensen} H. Esbensen et al., Phys. Rev. Lett. 41, 296 (1978).
\bibitem{Dasso} C. H. Dasso, Proc. of the la Rabida Int. Summer School on Nuclear Physics, p. 398, eds. M. Lozano and G. Madurga, World Scientific, Singapore, 1985. 
\bibitem{Guidry} M. W. Guidry et al., Phys. Rev. C 36, 609 (1987).  
\bibitem{Galetti} D. Galetti et al., Phys. Rev. C 48, 3131 (1993). 
\bibitem{DassoDonangelo} C. H. Dasso and R. Donangelo, Phys. Lett. B 276, 1 (1992). 
\bibitem{Esbensen4} H. Esbensen and S. Landowne, Phys. Rev. C 35, 2090 (1987).
\bibitem{Randrup2} J. Randrup, Nucl. Phys. A327, 490 (1979).
\bibitem{Feldmeier} H. Feldmeier, Rep. Prog. Phys. 50, 915 (1987).
\bibitem{Denisov} V. Yu. Denisov, Eur. Phys. J. A 7, 87 (2000). 
\bibitem{Birkelund} J. R. Birkelund et al., Phys. Rep. 56, 107 (1979).
\bibitem{Hasse} R. W. Hasse and W. D. Myers, Geometrical Relationships of Macroscopic Nuclear Physics, Springer-Verlag, Berlin, 1988.
\bibitem{Chamon} L. C. Chamon et al., Phys. Rev. C 66, 014610 (2002).
\bibitem{Ribeiro} M. A. C. Ribeiro et al., Phys. Rev. Lett. 78, 3270 (1997). 
\bibitem{Chamon2} L. C. Chamon et al., Phys. Rev. Lett. 79, 5218 (1997).  
\bibitem{Gasques} L. R. Gasques et al., Phys. Rev. C 69, 034603 (2004).
\bibitem{Esbensen2} H. Esbensen, Phys. Rev. C 77, 054608 (2008).
\bibitem{Esbensen3} H. Esbensen, C. L. Jiang, and K. E. Rehm, Phys. Rev. C 57, 2401 (1998).
\bibitem{Beckerman2} M. Beckerman, Phys. Rep. 129, 145 (1985). 
\bibitem{Nobre} G. P. A. Nobre et al., Nucl. Phys. A 786, 90 (2007). 
\bibitem{Jiang} C. L. Jiang et al., Phys. Rev. Lett. 93, 012701 (2004).
\bibitem{Ackermann} D. Ackermann et al., Nucl. Phys. A 609, 91 (1996).
\bibitem{Beckerman} M. Beckerman et al., Phys. Rev. C 25, 837 (1982).
\bibitem{Misicu} H. Misicu and H. Esbensen, Phys. Rev. Lett. 96, 112701 (2006); Phys. Rev. C 75, 034606 (2007).
\end{thebibliography}
\end{document}